\documentclass[aps,pra,showpacs,twocolumn]{revtex4}
\usepackage{graphicx}
\usepackage{bm}
\begin{document}
\preprint{}
\title{Alternative measures of uncertainty in quantum metrology: Contradictions 
and limits}
\author{Alfredo Luis}
\email{alluis@fis.ucm.es}
\homepage{http://www.ucm.es/info/gioq}
\author{Alfonso Rodil}
\affiliation{Departamento de \'{O}ptica, Facultad de Ciencias
F\'{\i}sicas, Universidad Complutense, 28040 Madrid, Spain }
\date{\today}

\begin{abstract}
We examine a family of intrinsic performance measures in terms of probability distributions
that generalize Hellinger distance and Fisher information. They are applied to quantum metrology 
to assess the uncertainty in the detection of minute changes of physical quantities. We  show that 
different measures lead to contradictory conclusions, including the possibility of arbitrarily 
small uncertainty for fixed resources. These intrinsic performances are compared with the averaged
error in the corresponding estimation problem after single-shot measurements.
\end{abstract}


\maketitle

\section{Introduction}

Quantum fluctuations and uncertainty are key issues in quantum physics because 
of the fundamental statistical nature of the theory. They also enter on practical 
matters such as quantum metrology,  where is of fundamental importance to
determine whether quantum fluctuations impose an ultimate lower limit to the uncertainty
in the detection of minute changes of physical quantities. The evidence supporting the 
universality of a lower bound known as Heisenberg limit \cite{Hl} is not as solid as it would 
be desirable as revealed by recent examples \cite{RL}. In some very recent approaches 
more conclusive results  are obtained by averaging uncertainty  over finite intervals for
the monitored variable representing our prior knowledge about it  \cite{amse}. 

Historically, the statistical inference about uncertainty is addressed by variance-based 
methods, mainly because they properly fit Gaussian statistics. Nevertheless, this may be 
not satisfactory enough in other situations, and alternative approaches may be of interest 
\cite{var,ypra,fase,atv,Ts,ML,L,ZZ}.  Previous works have already shown that 
different assessments of fluctuations may lead to contradictory and counterintuitive 
conclusions. For example, states with diverging variance may have arbitrary small entropy
for the very same observable \cite{ML}. This ambiguity extends to the uncertainty relation 
between complementary observables when using Renyi-Tsallis entropic measures, since
the very same state can be either of maximum or of minimum joint uncertainty, depending 
on the measure used  \cite{L}. Moreover, for some entropic uncertainty measures 
there is no lower bound to the joint uncertainty  of complementary observables \cite{ZZ} 
(see also Ref. \cite{HF}). 

After these precedents we think that it is worth investigating the application of alternative 
measures of uncertainty to the question of fundamental  resolution limits in quantum metrology
caused by quantum uncertainty. To this end we address an intrinsic evaluation of the 
detection performance  in terms of the closeness  between the probability distributions 
associated to  two close enough values of the monitored variable  \cite{DD}.  We use  
Renyi-Tsallis generalizations of the Hellinger distance and Fisher information.  
This is compared with the averaged error in the corresponding estimation problem.

\section{Generalized distances}

In general terms, a signal variable $\epsilon$ is monitored by the transformation 
$P \rightarrow P_\epsilon$ it induces in some observed quantity $P$. Within a 
quantum context we consider that $P = P(x)$ is the probability distribution of a 
given observable $X$, that will be assumed dimensionless, continuous, and unbounded, 
as a coordinate of a particle or a field quadrature, for example. Nevertheless, $P$ may 
equally well represent any other quantity in classical or quantum physics, such as the 
intensity of a light beam in classical optics, for example. 

We focus on the very usual case where the signal-induced transformation $P \rightarrow 
P_\epsilon$ is just a shift $P(x) \rightarrow P(x - \epsilon)$. The significance of $\epsilon$ 
can be assessed by the closeness between $P(x)$ and $P(x - \epsilon)$ \cite{NH}:
\begin{equation}
\label{gd}
D_q = \frac{1}{2} \int_{-\infty}^\infty dx \left | P^q (x-\epsilon) -
P^q (x) \right |^{1/q} ,
\end{equation}
where $q$ is any positive real number. These are Renyi-Tsallis generalizations of the
Hellinger distance, that is the case $q=1/2$ \cite{Ts}. We may regard $P^q$ as a kind 
of nonlinear record of $P$. This can be illustrated by the example of $P$ as light intensity
where  $P^q$ for integer $q>1$ may represent the nonlinear optical response of a 
nonlinear medium. On the other hand, $q=1/2$  is the case of homodyne detection, 
where the detector current is proportional to the amplitude of the signal electric field.

For weak signals $\epsilon \ll 1$ (the case of major interest in precision metrology) we have, 
to first order in $\epsilon$,
\begin{equation}
\label{Fq}
D_q \simeq \frac{q^{1/q}}{2} | \epsilon |^{1/q} F_q,  \qquad
F_q = \int_{-\infty}^\infty dx P(x) \left | \frac{d}{dx} \ln P(x) \right |^{1/q} .
\end{equation}
Therefore, $F_q$ is a family of generalizations of the Fisher information as  $1/q$-moments 
of the score, where the Fisher information is retrieved for $q=1/2$  \cite{Fi}. 

This provides an estimation of detection sensitivity as the minimum signal 
$\epsilon_\mathrm{min}$ required to exceed some threshold for $D_q$. For simplicity,
we assume that such threshold does not depend on the probe state, so that the sensitivity is 
solely determined by $F_q$ in the form
\begin{equation}
\label{ms}
\epsilon_{\mathrm{min}} = \frac{1}{F_q^q} .
\end{equation}

We have evaluated the generalized $F_q$ at  $\epsilon =0$. Nevertheless, the conclusions so 
obtained extend to other $\epsilon$ values because the signal-induced transformation is just 
a shift that does not affect the form of the probability distribution. This is at difference with other 
transformations, such as optical phase shifts, that change the form of the measured distribution 
for most probe states and practical observables, leading to Fisher information depending on 
the value of the signal. This is essentially the reason for averaging uncertainties over prescribed 
prior intervals for the monitored variable carried out in Ref. \cite{amse}.

In the general case $D_q^q$ is not a proper distance for $q \neq 1/2$, since the  triangle 
inequality may fail. Nevertheless, regarding metrological applications we are just interested
in the $\epsilon$-dependence around $\epsilon =0$, where, after Eq. (\ref{Fq}), $D_q^q \propto 
| \epsilon| F_q^q$, behaves as a proper distance.

\section{Probe state}

Let us consider probe states following an exponential power distribution (also referred to 
as generalized normal distribution, or generalized error distribution) illustrated in Fig. 1
\begin{equation}
\label{Px}
P(x) = \frac{\alpha 2^{1/\alpha}}{2 \gamma \Gamma (1/\alpha)} \exp 
\left ( - 2 \left | x/ \gamma \right |^\alpha \right ),
\end{equation}
where $\alpha$ and $\gamma$ are real nonnegative parameters. For $\alpha = 1$  this is the 
bound state of delta potentials $V(x) \propto - \delta (x)$, while for $\alpha = 2$ these are 
Gaussians including the fundamental state of harmonic oscillators, $V(x) \propto x^2$. On 
the other hand, for $\alpha \rightarrow \infty$ $P(x)$ tends to be a square distribution,
that for a free particle may be implemented with suitably arranged shutters acting on a momentum 
eigenstate  \cite{MM}. We assume that the signal-dependent transformation and the measurement 
are fast enough so we do not have to consider the free evolution of the probe during the process.

\begin{figure}
\begin{center}
\includegraphics[width=6cm]{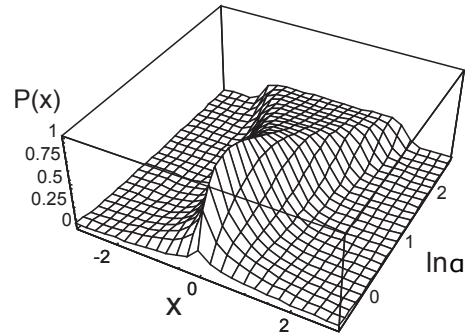}
\end{center}
\caption{Three-dimensional plot of $P(x)$ in Eq. (\ref{Px}) as a function of $x$ 
and $\ln \alpha$ with $\gamma$ given by Eq. (\ref{gE}) with $\langle E \rangle =1$.
Throughout this work plotted quantities are dimensionless.}
\end{figure}

The parameter $\gamma$ may be conveniently expressed in terms of the mean 
value of the energy $\langle E \rangle$, according to the usual practice 
in quantum metrology of relating resolution with the energy resources 
employed \cite{Hl,RL,amse}. For simplicity, let us assume that the probe is a 
free particle $E=p^2$, where $p$ represents the dimensionless variable 
canonically conjugate to $x$. From now on we consider the 
energetically optimum case where the wave-function is real, $\psi(x) = \sqrt{P(x)}$, 
so that $\langle p \rangle =0$. Then, the computation of $\langle E \rangle = \left ( 
\Delta p \right )^2 = \langle p^2 \rangle$ readily provides a relation between 
$\gamma$ and the mean energy $\langle E \rangle$ valid for $\alpha >1/2$
\begin{equation}
\label{gE}
\gamma = \frac{\alpha 2^{1/\alpha}}{2 \sqrt{\langle E \rangle}} 
\sqrt{\frac{\Gamma (2-1/\alpha)}{\Gamma (1/\alpha)}}. 
\end{equation}
The case $\langle p \rangle =0$ is energetically optimum because the 
sensitivity to $x-$shifts will depend on $\Delta x$ and $\Delta p$, but 
not on $\langle x \rangle$ or $\langle p \rangle$. Thus, the condition 
$\langle p \rangle =0$ avoids expending energy resources on dynamical 
states not related with the efficiency of the detection.º

\section{Alternative measures}

Next, the intrinsic performance assessment in Eq. (\ref{ms}) is compared with the 
estimation uncertainty after the measurement of $X$.  In all cases we follow the standard 
practice in quantum metrology of analyzing  detection performances for fixed energy 
resources \cite{Hl,RL,amse}.  We focus on single-shot measurement since it is 
energetically optimum to concentrate all resources in a single measurement \cite{Hl}.

\subsection{ Generalization of Fisher information}

By direct computation of $F_q$ after Eqs. (\ref{Fq}) and (\ref{Px}) we get 
for $\alpha > \max (1- q, 1/2)$
\begin{equation}
\label{FE}
F_q = \frac{\Gamma \left ( \frac{\alpha + q-1}{\alpha q} \right )}
{\Gamma (1/\alpha)} \left ( \frac{\alpha 2^{1/\alpha}}{\gamma} \right )^{1/q} . 
\end{equation}
After Eqs. (\ref{ms}), (\ref{gE})  and (\ref{FE}) we get
\begin{equation}
\label{eps}
\epsilon_{\mathrm{min}} = \frac{\Gamma^{q} \left ( 1/\alpha \right )}{
\alpha 2^{1/\alpha} \Gamma^q \left ( \frac{\alpha + q -1}{q \alpha } \right )}  \gamma.
\end{equation}
In Fig. 2 we have plotted $\epsilon_{\mathrm{min}}$ as a function of $\alpha$ for several 
values of $q$. We can appreciate strong differences arising for different values of $q$ 
and $\alpha$. For $q>1/2$ we have $\epsilon_\mathrm{min} \rightarrow \infty$ both for 
$\alpha \rightarrow 1/2$ and $\alpha \rightarrow \infty$, being $\epsilon_\mathrm{min}$ 
minimum for $\alpha =1$. On the other hand, for $q<1/2$ we have the exact opposite 
behavior with $\epsilon_\mathrm{min} \rightarrow 0$ both for $\alpha \rightarrow 1-q$ 
and $\alpha \rightarrow \infty$, providing $\alpha =1$ the maximum for 
$\epsilon_\mathrm{min}$. Finally, for $q=1/2$ there is no dependence on $\alpha$.

\begin{figure}
\begin{center}
\includegraphics[width=6cm]{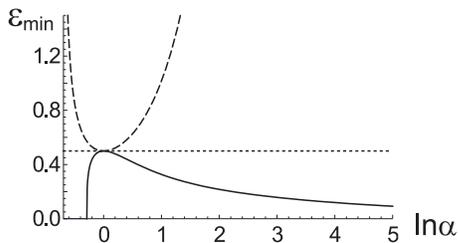}
\end{center}
\caption{Plot of $\epsilon_{\mathrm{min}} $ for fixed $\langle E \rangle=1$
as a function of $\ln \alpha$ for $q =1/4$ (solid line) $q=1/2$ (dotted 
line), and $q= 2$ (dashed line). }
\end{figure}

\subsection{Width of posterior distribution}

From a Bayesian perspective, after any outcome $x$ we can infer a conditional probability 
distribution for the estimate $\tilde{\epsilon}$  of $\epsilon$ as $P(\tilde{\epsilon} |x) = 
P(x-\tilde{\epsilon})$. A suitable measure of the estimation uncertainty is given by the width of  
$P(\tilde{\epsilon} |x)$ as a function of  $\tilde{\epsilon}$. Following the spirit of the preceding sections 
we may consider  Renyi-Tsallis  measures of uncertainty $\delta \tilde{\epsilon}_q$ (or generalized
Fisher lengths)  as \cite{HR,HF} :
\begin{equation}
\label{Lq}
\delta \tilde{\epsilon}_q = \left [ \int dx P^q (\tilde{\epsilon} | x ) \right ]^\frac{1}{1-q}  = 
\frac{1}{q^{\frac{1}{\alpha (1-q)}}} \frac{2 \Gamma (1/\alpha)}{\alpha 2^{1/\alpha} }  \gamma ,
\end{equation}
which are independent of $x$. In Fig. 3 we have represented $\delta \tilde{\epsilon}_q $ 
showing that it is quite similar for all values of $q$ examined, and also very similar  to the case 
$q>1/2$ of $\epsilon_{\mathrm{min}}$.

\begin{figure}
\begin{center}
\includegraphics[width=6cm]{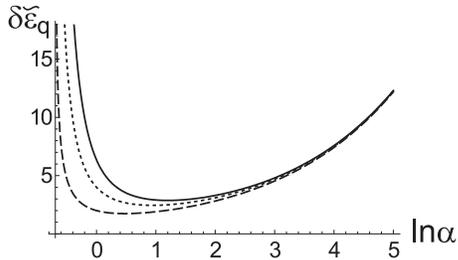}
\end{center}
\caption{Plot of $\delta \tilde{\epsilon}_q $ in Eq. (\ref{Lq})  as function of $\alpha$ for  fixed 
$\langle E \rangle =1$ and $q =1/4$ (solid line), $q=1/2$ (dotted line), and $q= 2$ (dashed line).}
\end{figure}

\subsection{Mean estimation error}

After a single observation the outcome $x$ is a suitable unbiased estimator of $\epsilon$ since its 
average coincides with the true value $\int dx  x P(x|\epsilon) = \epsilon$, where  $P(x|\epsilon) = 
P(x- \epsilon)$ is the probability of  $x$ conditioned to $\epsilon$.   In the same spirit of the above 
generalization we may consider the mean estimation error
\begin{equation}
\label{me}
\Delta \epsilon_q = \left [ \int dx P(x|\epsilon) | x-\epsilon |^{1/q} \right ]^q =
\frac{1}{2^{1/\alpha}}  \frac{\Gamma^q \left ( \frac{1+q}{\alpha q} \right )}{\Gamma^q (1/\alpha) } 
\gamma,
 \end{equation}
which is independent of $\epsilon$. In Fig. 4 we have  represented $\Delta \epsilon_q $  showing 
that it is quite  similar to $\delta \tilde{\epsilon}_q $ for all values of $q$ examined.

\begin{figure}
\begin{center}
\includegraphics[width=6cm]{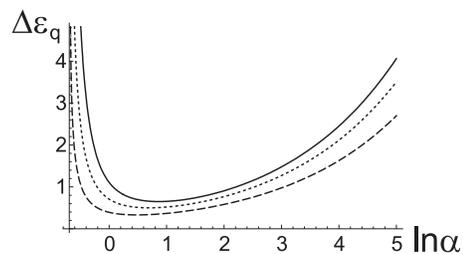}
\end{center}
\caption{Plot of  $\Delta \epsilon_q $ in Eq. (\ref{me}) as function of $\alpha$ for  fixed $\langle E
 \rangle =1$ and $q =1/4$ (solid line), $q=1/2$ (dotted line), and $q= 2$ (dashed line).}
\end{figure}
\section{Discussion and conclusions}

We have examined the  assessment of uncertainty provided by a family of generalizations 
of the Hellinger distance and the Fisher information. After considering probes in exponential power 
distributions we have obtained two relevant conclusions: 

(i) Different measures lead to contradictory conclusions. This is illustrated by the cases $q=1/4$ 
and $q=2$ in Fig. 2 where $\alpha=1$ provides the maximum sensitivity for $q >1/2$ and the 
minimum sensitivity for $q <1/2$. 

(ii) For  $q <1/2$ we get increasingly high sensitivity for probes with properly chosen values of 
$\alpha$ for fixed and finite resources. 

We have computed also similarly generalized averaged errors in the corresponding single-shot 
estimation problem, showing that they do not reproduce the two above features. Thus we wonder 
whether there is any suitable counterpart of the Cramer-Rao lower bound involving these generalized 
performance measures.

We think that this approach  may provide useful insights for the understanding of 
uncertainty, uncertainty relations, and their implications in quantum metrology. The  vanishing 
of $ \epsilon_{\mathrm{min}}$ for finite energy resources that arise for  $q<1/2$ seems 
to defy basic ideas in quantum metrology about ultimate resolution limits. It would be of interest
to determine whether this result is related to the lack of lower bound to the joint uncertainty 
of incompatible observables that arises for related Renyi-Tsallis uncertainty measures
 \cite{ZZ} .  On the other hand, in previous works we have found that the relation between 
 quantum limits and uncertainty relations is not trivial since probe states leading to minimum 
 metrological uncertainty may be far from being minimum uncertainty states \cite{mus}.
 
The result (ii) also recalls  the increasing sensitivity of Fabry-Perot arrangements for 
increasing mirror reflectivity \cite{FP}. This might be regarded as a kind of instrumental 
factor provided by the probe state that can be exploited for some values of $q$.
Moreover,  the ambiguity reported in conclusion (i) might be solved if physical 
reasons might impose that the performance measures used should be adapted to the 
probe state considered in each case.

\section*{ACKNOWLEDGMENTS}

A. L. acknowledges support from Projects No. FIS2012-35583 of the Spanish Ministerio de 
Econom\'{\i}a y Competitividad, and QUITEMAD S2009-ESP-1594 of the Comunidad de Madrid.


\begin{thebibliography}{00}

\bibitem{Hl} 
V. Giovannetti, S. Lloyd, and L. Maccone,Phys. Rev. Lett. \textbf{96}, 010401 (2006);
M. Zwierz, C. A. Perez-Delgado and P. Kok,\textit{ibid.} \textbf{105}, 180402 (2010);
V. Giovannetti, S. Lloyd, and L. Maccone, Science \textbf{306}, 1330 (2004);

\bibitem{RL}
A. Rivas and A. Luis, New J. Phys. \textbf{14}, 093052 (2012).

\bibitem{amse}
M. Tsang, Phys. Rev. Lett. \textbf{108}, 230401 (2012); 
V. Giovannetti and L. Maccone, \textit{ibid.} \textbf{108}, 210404 (2012);
Y. Gao and H. Lee, J. Phys. A \textbf{45}, 415306 (2012);
V. Giovannetti, S. Lloyd, and L. Maccone, Phys. Rev. Lett. \textbf{108}, 260405 (2012);
M. J. W. Hall, D. W. Berry, M. Zwierz, and H. M. Wiseman, Phys. Rev. A \textbf{85}, 041802(R) (2012);
M. J. W. Hall and H. M. Wiseman, New J. Phys. \textbf{14}, 033040 (2012);
R. Nair, arXiv:1204.3761v1.

\bibitem{var}
J. Hilgevoord, Am. J. Phys. \textbf{70}, 983 (2002); 
G. N. Lawrence, Laser Focus World \textbf{30}, 109 (1994);
J. \v{R}eh\`{a}\v{c}ek and Z. Hradil, J. Mod. Opt. \textbf{51}, 979 (2004).

\bibitem{ypra}
A. Luis, Phys. Rev. A \textbf{64}, 012103 (2001); 
\textbf{67}, 032108 (2003).

\bibitem{fase}
J. M. L\'{e}vy-Leblond, Ann. Phys. (N.Y.)\textbf{101}, 319 (1976);
E. Breitenberger, Found. Phys. \textbf{15}, 353 (1985);
J. B. M. Uffink, Phys. Lett. A \textbf{108}, 59 (1985);
J. M. L\'{e}vy-Leblond, \textit{ibid.} \textbf{111}, 353 (1985);
S. M. Barnett and D. T. Pegg, J. Mod. Opt. \textbf{36},7 (1989);
Z. Hradil, Phys. Rev. A \textbf{46}, R2217 (1992); 
Quantum Opt. \textbf{4}, 93 (1992); 
T. Opatrn\'{y}, J. Phys. A \textbf{27}, 7201 (1994);
V. Pe\v{r}inov\'{a}, A. Luk\v{s}, and J. Pe\v{r}ina,
\textit{Phase in Optics} (World Scientific, Singapore, 1998).

\bibitem{atv}
I. I. Hirschman, Am. J. Math. \textbf{79}, 152 (1957);
I. Bialynicki-Birula and J. Mycielski, Commun. Math. Phys. \textbf{44}, 129 (1975);
D. Deutsch, Phys. Rev. Lett. \textbf{50}, 631 (1983);
M. H. Partovi, \textit{ibid.} \textbf{50}, 1883 (1983); 
K. Kraus, Phys. Rev. D \textbf{35}, 3070 (1987);
H. Maassen and J.B.M. Uffink, Phys. Rev. Lett. \textbf{60}, 1103 (1988);
J. Sanchez, Phys. Lett. A \textbf{173}, 233 (1993); 
\v{C}. Brukner and A. Zeilinger, Phys. Rev. Lett. \textbf{83}, 3354 (1999);
Phys. Rev. A \textbf{63}, 022113 (2001);
V. Majern\'{\i}k and E. Majerníkov\'{a}, Rep. Math. Phys. \textbf{47}, 381 (2001);
M. J. W. Hall, Phys. Rev. A \textbf{64}, 052103 (2001);
S. Massar and P. Spindel, Phys. Rev. Lett. \textbf{100}, 190401 (2008); 
S. Wehner and A. Winter, New J. Phys. \textbf{12}, 025009 (2010); 
I. Bialynicki-Birula and L. Rudnicki arXiv:1001.4668v1;
I. Urizar-Lanz and G. T\'{o}th, Phys. Rev. A \textbf{81}, 052108 (2010); 
A. D. C. Nascimento, R. J. Cintra, and A. C. Frery, IEEE Trans. Geos. 
Remot. Sens. \textbf{48}, 373 (2010);
P. S\'{a}nchez-Moreno, A. R. Plastino, and J. S. Dehesa, J. Phys. A \textbf{44},
065301 (2011);
A. E. Rastegin, \textit{ibid.} \textbf{44}, 095303 (2011).  

\bibitem{Ts}
A. Renyi, "On the measures of entropy and information," in
Proceedings of the 4th Berkeley Symposium on Mathematics
and Staticstical Probability (University of California Press,
Berkeley, 1961), Vol. 1, pp. 547–561;
C. Tsallis, J. Stat. Phys. \textbf{52}, 479 (1988);
U Larsen, J. Phys. A \textbf{23}, 1041 (1990);
E.M.F. Curado and C. Tsallis, \textit{ibid.} \textbf{24}, L69 (1991);
A. K. Rajagopal, Phys. Lett. A \textbf{205}, 32 (1995);
I. Csisz\'{a}r, IEEE Trans. Inf. Theory \textbf{41}, 26 (1995);
M. Portesi and A. Plastino, Physica A \textbf{225}, 412 (1996); 
I. Bialynicki-Birula, Phys. Rev. A \textbf{74}, 052101 (2006);
F. Chapeau-Blondeau, A. Delahaies, and D. Rousseau, Phys. Lett. A 
\textbf{375}, 2211 (2011).

\bibitem{ML}
P. Mat\'{\i}a-Hernando and A. Luis, Phys. Rev. A \textbf{86}, 052106 (2012).

\bibitem{L}
A. Luis, Phys. Rev. A \textbf{84}, 034101 (2011).

\bibitem{ZZ}
M. Zakai, Inf. Control \textbf{3}, 101 (1960);
A. Luis, Opt. Lett. \textbf{31}, 3644 (2006);
Phys. Rev. A \textbf{75}, 052115 (2007);
S. Zozor, M. Portesi, and C. Vignat,  Physica A \textbf{387}, 4800 (2008).

\bibitem{HF}
M. J. W. Hall, Phys. Rev. A \textbf{62}, 012107 (2000).

\bibitem{DD}
Z. Hradil, R. My\v{s}ka, T. Opatrn\'{y}, and J. Bajer, 
Phys. Rev. A \textbf{53}, 3738 (1996);
G. A. Durkin and J. P. Dowling, Phys. Rev. Lett. \textbf{99}, 
070801 (2007). 

\bibitem{NH}
N. Hadjisavvas, Linear Algr. Appl. \textbf{84}, 281 (1986);
Z.-H. Ma and J.-L. Chen, J. Phys. A \textbf{44}, 195303 (2011).

\bibitem{Fi}
H. Cram\'{e}r,  {\it Mathematical methods of statistics,} (Asia Publishing House, Bombay, 1962);
T. M. Cover and J. A. Thomas, {\it Elements of Information Theory,}
(Wiley Interscience, New York, 1991);
C. W. Helstrom, Phys. Lett. A \textbf{25}, 101 (1967);
IEEE Trans. Inf. Theory \textbf{IT-14}, 234 (1968);
\textit{Quantum Detection and Estimation Theory}
(AcademicPress, New York, 1976);
S. L. Braunstein and C. M. Caves, Phys. Rev. Lett. \textbf{72},
3439 (1994).

\bibitem{MM}
M. Moshinsky, Phys. Rev. \textbf{88}, 625 (1952); 
E. Torrontegui, J. Mu\~{n}oz, Y. Ban, and J. G. Muga, 
Phys. Rev. A \textbf{83}, 043608 (2011).

\bibitem{HR}
M. J. W. Hall, Phys. Rev. A \textbf{59}, 2602 (1999).

\bibitem{mus}
D. Maldonado-Mundo and A. Luis, Phys. Rev. A \textbf{80}, 063811(2009);
A. Luis, Ann. Phys. (N. Y.) \textbf{331}, 1 (2013).

\bibitem{FP}
A. Luis and L. L. S\'{a}nchez-Soto, J. Mod. Opt. \textbf{38}, 971 (1991).

 
\end{thebibliography}
\end{document}